# Blockchain Signatures to Ensure Information Integrity and Non-Repudiation in the Digital Era: A comprehensive study

Kaveri Banerjee[1], Sajal Saha[2]
[1]Research Scholar, Department of CSE, SOET, Adamas University, Kolkata,
Senior Assistant Professor & Head of Dept BCA, Nopany Institute of Management Studies, Kolkata
[2]Prof. & Head of Dept CSE, SOET, Director Production and Innovation, Adamas University, Kolkata
[1]kg.kaveri7@gmail.com, [2]sajalkrsaha@gmail.com

**Abstract**

Blockchain, recognized for its revolutionary potential, has gained significant attention because of its key attributes, such as a decentralized ledger and robust security. Among these attributes, non-repudiation is a critical part of information security within blockchain systems. Ensuring non-repudiation can be effectively accomplished using digital signature schemes. The purpose of this study is to analyze digital signatures used in blockchain technology, focusing on their role in achieving non-repudiation and preserving information integrity. This study addresses the need for non-repudiation in blockchain systems to prevent denial of transaction authenticity and maintain information integrity. Using exploratory digital signature schemes, the research aimed to understand how these cryptographic tools contribute to non-repudiation as well as overall security in blockchain applications. This research involves an extensive exploration of blockchain features, particularly the role of digital signatures in maintaining information integrity. A comprehensive analysis of various digital signature schemes is conducted, encompassing their mathematical foundations, security assumptions, as well as cryptographic properties. This study emphasizes the need for digital signatures to ensure non-repudiation in blockchain systems. This section presents a complete overview of several digital signature algorithms, assessing each based on the aforementioned cryptographic features. This study highlights each scheme's benefits and disadvantages in the context of blockchain technology.

**Keyword** Digital Signature, Bitcoin, Internet of Things, RSA, blockchain technology

## 1. Introduction

Blockchain technology, presently a cornerstone of Bitcoin, has received extensive research and implementation [1]. It offers a decentralized framework across various domains [2] and maintains an unbroken, tamper-resistant chronological data ledger [3]. Moreover, it has become a major subject of research interest [4]. Blockchain is an ideal platform for asserting ownership [5] and providing timestamped digital evidence when physical or digital assets can be represented as digital digests [6].

Consequently, blockchain technology is applicable across various industries [7], spanning online transactions, stock markets, trade administration, and the Internet of Things (IoT) [8]. This innovation serves as an underpinning not just for digital currencies [9] but also for broader progress and applications within conventional finance as well as trade [10]. Moreover, it introduces innovative ideas like smart contracts [11]. The growth of blockchain technology has brought to the forefront issues like hard forks and double payments, primarily stemming from conflicts within blocks [12]. These concerns directly affect the validity and trustworthiness of transactions in E-commerce. At their core, these issues revolve around information security, encompassing various aspects such as the resilience of consensus processes against 51% of attacks [13, 14], vulnerabilities to Byzantine failures, potential vulnerabilities in asymmetric encryption algorithms, external threats and attacks, transaction data forgery, individual privacy breaches, and software design faults.

Previous studies have identified several limitations of blockchain-based digital signatures, such as scalability issues due to blockchain size, robustness inefficiencies in consensus mechanisms such as proof of work, legal uncertainty over the legal integrity of blockchain signatures, privacy concerns from the public nature of blockchain transactions, and challenges in maintaining specialized storage and secure methods of management. The study aims to investigate digital signatures used in blockchain technology to ensure non-repudiation and information integrity. Specific objectives include investigating various digital signature schemes to better understand their mathematical foundations, security assumptions, and cryptographic properties; evaluating these schemes using critical cryptographic properties; identifying their strengths and weaknesses within blockchain systems; and proposing future research directions to improve digital signature algorithms for blockchain applications. Digital signatures are critical for ensuring transaction authenticity (non-repudiation), maintaining data integrity, authenticating participant identities, protecting against tampering and fraud (security), and enabling large-scale adoption and high performance. By thoroughly analysing these factors, the study hopes to enhance secure and efficient blockchain applications, hence increasing the technology's robustness and reliability across a variety of industries.

A detailed review of digital signature schemes in the blockchain area shows the various approaches, security measures, and performance aspects they present. ECDSA (Elliptic Curve Digital Signature Algorithm), generally used for authentication in networks including Bitcoin and Ethereum, ensures strong security through elliptic curve cryptography at the expense of computational cost. RSA (Rivest-

Shamir-Adleman), which is based on the factorization of primes, furnish strength but is slow and generates longer signatures. The EdDSA (Elliptic Curve Digital Signature Algorithm), based on efficient elliptic curve implementations, has faster operations and better protection against side-channel attacks than the previous algorithms. Schnorr signatures (Boneh-Lynn-Shacham) widely known for their simplicity, stand out for provable security, particularly in the efficiency, more so in batch verification scenarios. Based on bilinear pairings, BLS signatures facilitate efficient batching and provide the security guarantees. These schemes provide a choice between several factors, like security requirements, performance concerns, and compatibility with blockchain protocols, among which ongoing research is expected to improve security, efficiency and scalability in applications of blockchain

Previous research on digital signatures and blockchain technologies revealed severe flaws. There needs to be a more comprehensive examination of all digital signature schemes in the context of blockchain, with research frequently focused on specific schemes while ignoring concerns such as security, performance, and scalability. Furthermore, there needs to be more study into creative digital signature algorithms designed for blockchain optimization, which has hampered scheme adaptation and development for blockchain applications. Despite interest in newer and more powerful digital signature algorithms for blockchain, the present research focuses primarily on proven schemes such as ECDSA and RSA. Furthermore, there needs to be more research into using digital signatures for applications other than transaction authorization, such as smart contracts. The interaction between digital signatures and other blockchain aspects, such as consensus techniques, privacy, and storage, must be better understood. Finally, there needs to be more forward-looking research that will develop digital signatures to satisfy future blockchain requirements, such as scalability and new decentralized designs. Covering these challenges in a subsequent research cycle can significantly improve blockchain-based systems' security, performance, and innovation.

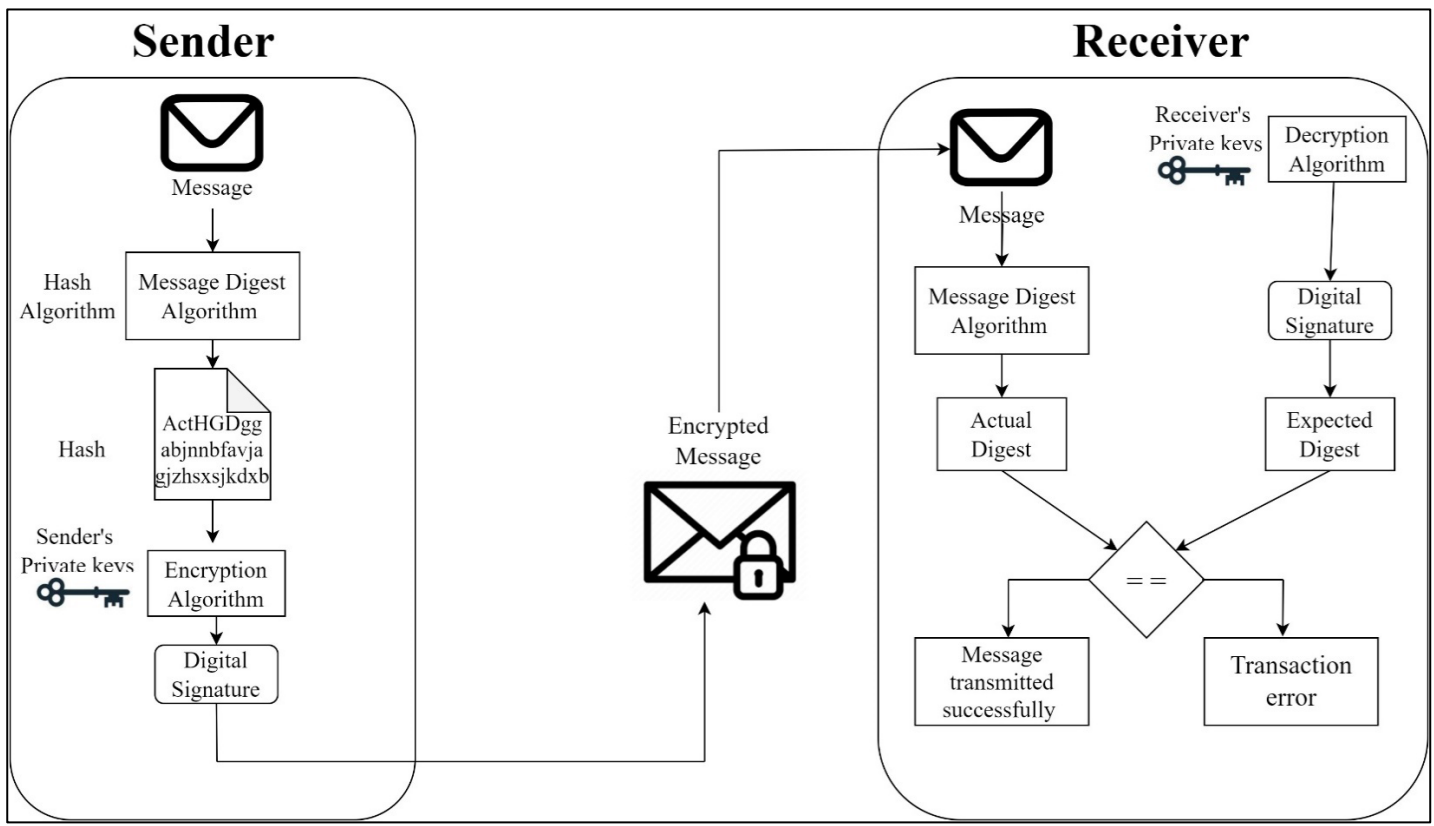

Figure 1: Fundamentals of Data Security Using RSA-Based Digital Signature

Establishing information non-repudiation is vital to the security of blockchain. This guarantee is made possible by several security solutions incorporated within the blockchain system. They involve the use of digital signatures and their exploration of identity authentication methods, timestamping techniques, etc. Digital signatures are important in maintaining the authenticity of data or messages and have a feature known as non-repudiation. Figure 1 uses the RSA algorithm, a frequently used technique for digital signatures. We provide this as an example of a typical digital signature process. Digital signature technology is very suited to the intrinsic features of a blockchain system. In this environment, it enhances security and increases its possible applications beyond old ones. It adds tremendous value to the process. In older environments, digital signatures authorize data on the Internet; they do not imply value exchange. The signatures can be applied more safely and widely in the blockchain, providing room for value creation.

## 1.1. Research Contributions:

We are presented with the transfer of real value as well as the transmission of data within the framework of a blockchain system. Our paper makes the following crucial contributions in this regard:

1. We systematically review, analyze, and compare digital signature schemes to ensure information non-repudiation within blockchain technology. Additionally, we delve into the latest advancements in digital signatures. This comprehensive examination aims to streamline the design and optimization of signature algorithms for enhanced performance and security.
2. We suggest new lines of inquiry regarding the digital signatures and offer prospective routes for further investigation.

### 1.2. Research Gaps:

After reviewing the literature on digital signatures in the blockchain domain, several potential literature gaps can be identified:

- Understanding the integration challenges with emerging blockchain technologies.
- Comprehensive analyses of scalability and performance trade-offs of virtual signature schemes.
- Research on real-world implementation challenges and adoption barriers.
- Exploration of security and privacy trade-offs specific to blockchain applications.
- Understanding interoperability challenges with traditional systems.
- Addressing regulatory and legal implications of virtual signatures in blockchain applications.

### 1.3. Research Questions:

- RQ1: How can digital signature schemes undergo systematic review, analysis, and comparison to ensure information non-repudiation within blockchain technology?
- RQ2: What are the latest expansions in digital signatures, and how can they improve signature algorithms for better performance and security?
- RQ3: What are the recent research directions for digital signatures in blockchain?

The succeeding sections of this paper are organized in the following manner: Section 2 commences a discourse on blockchain technology and its interaction with non-repudiation. In Section 3, we discuss the research methodology. Section 4 meticulously examines various digital signature schemes. Building on this, Section 5 delves into an extensive review and comparative analysis of pertinent works within this field. In Section 6, we discuss the results. Finally, in Section 7, we conclude our paper while outlining prospective avenues for future research.

## 2. Blockchain and the concept of non-repudiation

### 2.1 Blockchain

Nakamoto initially proposed the blockchain concept in 2008, and it has for the reason that ends it ends up being the underlying technology underpinning Bitcoin. Initially used one at a time, the terms "block" as well as "chain" were later combined to produce the famous term "blockchain." A "block" in the blockchain refers to a recorded issue of Bitcoin transaction facts and comprises the block header and the block content material, which can be vital elements. The period "blockchain" wasn't formally recognized until 2016 [15]. Blockchains may be divided into subsequent classes in a broader feel: Public Blockchain, Consortium Blockchain, and Private Blockchain. A network can also be created by connecting numerous blockchains, and the linkages and chains that make up the network shape an interchain device [13]. Figure 2 indicates how blockchain relationships are categorized.

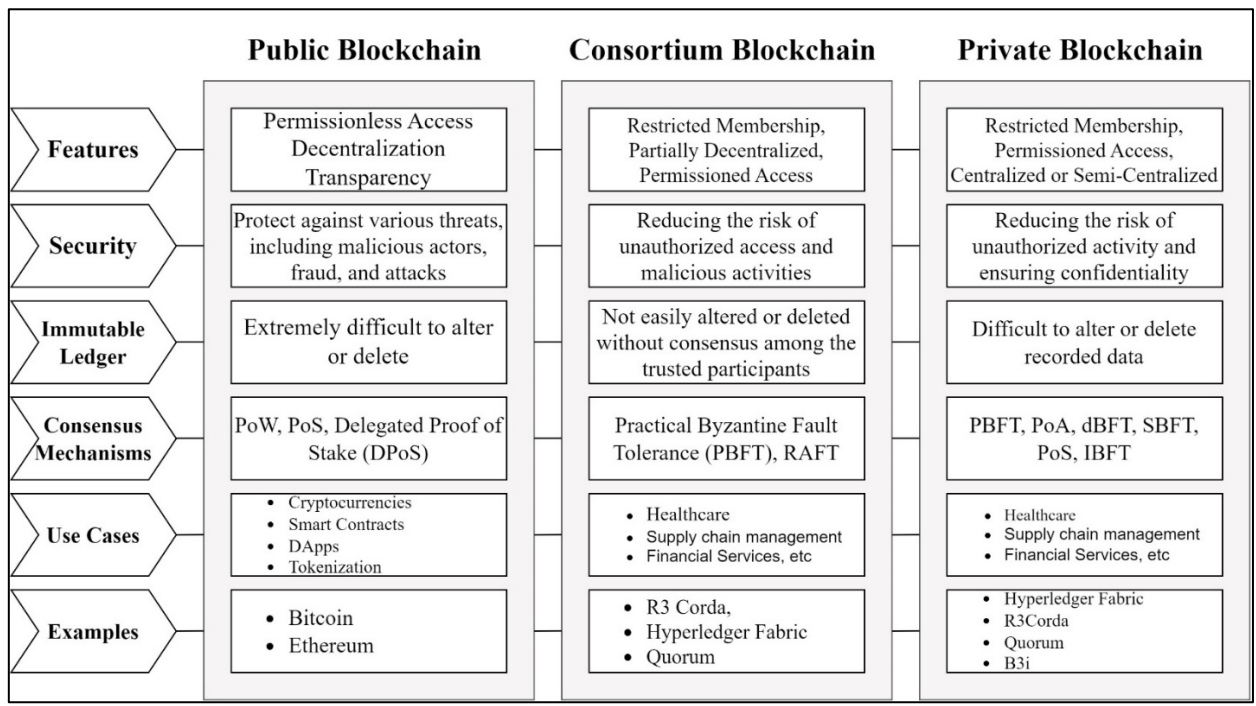

Figure 2: Categories of Blockchain

There are also a few specific characteristics of the blockchain era, which include decentralization, transparency, self-governance and immutability [16]. Some of its simple elements are, for instance, uneven encryption, peer-to-peer (P2P) conversation, dispensed ledger structure, consensus procedures, and clever contracts. Security techniques and algorithms are also a part of blockchain systems [17]. Developing facts technology, including Cloud Computing, the Internet of Things (IoT) [18], and massive information, are gradually synergizing with blockchain technology [19]. These technologies provide the vital foundations of contemporary infrastructures. As a stimulus to the development of next-era records generation, it enables a higher-stage safety standard regardless of mass market Internet things.

In addition, blockchain and IoT were intently incorporated by researchers [20]. This integration promotes provider and aid sharing, units up marketplaces for services among IoT gadgets streamlines encryption and authentication in complex workflows. According to investigative findings [21], combining blockchain and IoT has a big capacity for innovating across industries, generating new business patterns, and even giving an upward push to distributed requests.

### 2.2 Non-Repudiation
Non-repudiation guarantees that applicants in blockchain-based E-commerce transactions cannot dispute their involvement or the activities recorded on the blockchain. Non-repudiation services' main goal is to assemble, store, deliver, and authenticate irrefutable evidence regarding messages sent between the sender and receiver. A delivery authority (DA), often known as a reliable third party, may be required in this situation [22].
In blockchain technology, non-repudiation has two crucial aspects:
1. Sender's Non-Repudiation: This dimension ensures that the sender of information cannot deny their actions. For instance, if owner A sends a message to owner B, owner A cannot later deny this transaction.
2. Receiver's Non-Repudiation: This facet ensures that the receiver of information cannot deny receiving it. In other words, if owner A sends a message to owner B, owner B cannot deny receipt of the message.

To achieve non-repudiation in blockchain systems, digital signatures are employed, using asymmetric encryption techniques, often based on elliptic curve equations [23]. These digital signatures provide a cryptographic guarantee of information non-repudiation.
As an example, elliptic curves and modular arithmetic in finite fields are used to create digital signatures in the context of Bitcoin [24]. Since the message's sender needs to possess the private key to prove ownership of the associated Bitcoins, these signatures offer non-repudiation. Any network participant can confirm this ownership and the validity of the transaction, as illustrated in Figure 3.

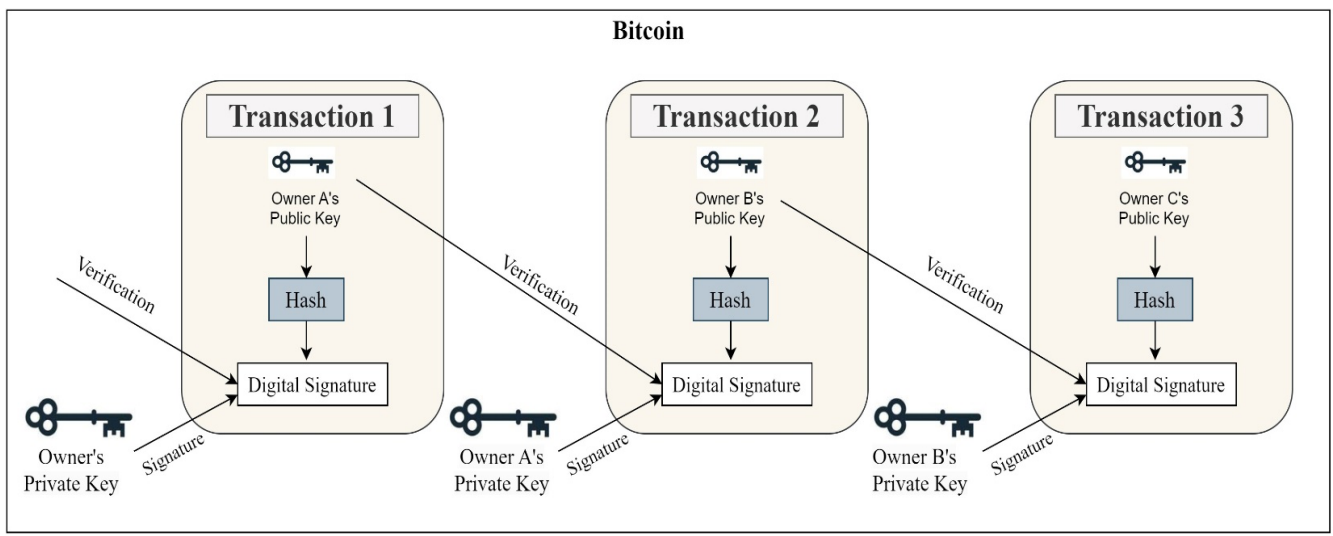

Figure 3: The Digital Signature Process in Bitcoin Transactions

## 3. Methodology

The paper outlines several key criteria for selecting and evaluating digital signature schemes within the context of blockchain technology:

- Non-repudiation: The signature scheme's capacity to offer convincing evidence that the signer could not deny signing the message. This is critical to ensure the validity and integrity of blockchain transactions.
- Data integrity: The ability of the signature system to detect tampering or change of signed data, hence protecting data integrity.
- Security assumptions and cryptographic properties: The paper analyzes the mathematical underpinnings, security assumptions, and cryptographic attributes like unforgeability, blindness, anonymity, unlinkability, etc. that form the basis of the different digital signature schemes.
- Performance metrics: Evaluation of the computational efficiency, signature sizes, verification times, and other performance aspects of the schemes, which impact their authentication, integrity, non-repudiation, as well as other important aspects like efficiency, security properties, scalability, and long-term viability of the schemes across different applications.
- Compatibility with blockchain protocols: The ability of the signature schemes to seamlessly integrate with existing blockchain frameworks like Bitcoin and Ethereum.
- Application scenarios: The suitability of the signature schemes for different blockchain use cases, such as cryptocurrency transactions, smart contracts, and the Internet of Things, based on their security and functional properties.

A comparison and contrast of the strengths, weaknesses, and trade-offs of different signature schemes, including aggregate, group, ring, blind, and proxy signatures, when applied to blockchain systems.
The evaluation aims to identify signature algorithms that can effectively address the unique security, performance, and functional requirements of blockchain environments while ensuring robust non-repudiation and information integrity guarantees.

## 4. Exploring Digital Signature Techniques in Blockchain

A digital signature is a form of cryptography verifying that documents or messages are authentic and in their original state. The private key creates a unique digital "signature" for every data piece. It may then be checked with the corresponding public key. Blockchain technology is one of the many secure communication and data verification technologies that depend heavily on digital certificates [25]. They are a form of non-repudiation where they help affirm that an exchange or transaction over the network has not undergone any changes and was sent by its purported originator.

This is a detailed overview and an evaluation of existing well-established digital signature techniques often used throughout the blockchain industry. The types of signed procedures under investigation include group signatures, combined signatures, blinded signature schemes, ring names and proxy signing.

### 4.1. Aggregate Signatures: Streamlining Multi-Signature Verification

A separate class of digital signature schemes is aggregate signatures. These signatures include an accumulation function, generally based on concepts such as co-GDH (Computational Diffie-Hellman) and bilinear mapping [26]. The basic purpose of an aggregated signature is to gather many individual signatures- each representing a different message contributed by the users, into one concise and orderly name.

Although concise, this combined signature effectively tells verifiers that all users taking part have signed their respective messages. Put, the aggregate signatures make it much easy just to authenticate. When there are signals that need validating in large quantities at a time, this is essential. This feature makes aggregate signatures quite powerful, especially when efficiency and scalability are of the essence.

### 4.2. Group Signatures: Balancing Anonymity and Security

Group signature systems empower participants inside a collection to signal messages while keeping their identities collectively. These systems are constrained utilizing stringent standardized safety necessities, consisting of the following.

- Trust and Authenticity: Provides message integrity by way of ensuring signatures.
- Non-forgery privilege: To prevent unlicensed entities from counterfeiting signatures.
- Anonymity: To maintain the anonymity of the character individuals of a signed organization.
- Traceability: Enables legal entities to hint at the signature vicinity.
- Unlinkability: It makes it impossible to mix the same consumer's signatures into exceptional messages.
- No framing: Team contributors are prohibited from forging signatures and framing others.
- Non-invasive trace verification: Limits traceability wherein allowed.
- Anti-Integration: Protect in opposition to malicious companies of people colluding to check in underneath the maximum deceptive handshakes.

Considering many exceptional use cases for organization signature systems, their efforts are important regarding blockchain packages. The effectiveness of the group signature technique is typically evaluated by the usage of parameters, which include the period of the organization signature, the dimensions of the general public key, the time taken to execute the signature, and the period of the authentication manner.

### 4.3. Blind Signature: Safeguarding Privacy and Confidentiality

A blind signature [27] is a cryptographic approach that lets a person get a valid signature on a message at the same time as concealing the communique's authentic contents from the signer. The message is largely "blinded" or hidden in this scheme before it is submitted for signing, so the signer can't likely apprehend the message.

This signature scheme is a treasured tool for upholding privacy and confidentiality across various applications, such as digital coin systems and electronic voting. Blind signatures are designed to save you, the signer, from gaining any perception of the message's content while generating a legitimate signature. Blind signatures are essential in maintaining the privateness and security of virtual transactions and interactions, specifically in contexts where confidentiality holds paramount significance. In addition to meeting the typical requirements of virtual signatures, blind signatures adhere to the following vital factors:

1. Signer Invisibility: In a blind signature, the signer stays blind to the content material of the message they have recommended. This implies that the signer needs to gain expertise in the unique information or facts in the message for which they have furnished a signature.

2. Untraceability of Signed Message: Another fundamental asset is the untraceability of the signed message. Once the reduced size message is publicly disclosed or posted, the signatory cannot figure out which unique message they signed a few of the set of signed messages. This guarantees that the signing remains unlinkable to the signed content material, maintaining anonymity and privacy.

### 4.4. Ring Signature: Anonymity and Untraceability

The ring signature scheme uses the general public keys of all contributors within a certain set and a single personal key from a participant within that equal set [28]. This type of signature scheme is perfect for programs wherein keeping anonymity and ensuring untraceability is crucial, including anonymous payments or transactions requiring identity concealment.

A big function of a Ring signature scheme is its autonomy from a primary trusted entity. It ensures that signers keep anonymity, which benefits eventualities requiring long-lasting statistics protection. This scheme affords robust security, even in cases wherein an attacker (A) possesses the personal keys of all ring participants. In such instances, it cannot definitively decide the actual signer, with the probability of efficaciously identifying them being simplest $1/n$, wherein n represents the total number of ring contributors. Moreover, A faces substantial demanding situations in creating a legitimate ring signature for a message, substantially decreasing the chance of success.

Usually, a dependable ring signature should satisfy the fundamental protection stipulations:

- Unconditional Anonymity: The ring signature needs to offer absolute anonymity, making sure that even if an unauthorized attacker acquires the non-public keys of all ability signers, their capability to pinpoint the actual signer must not surpass an opportunity of $1/n$, in which n denotes the total count of capability signers. Essentially, the real signer must stay surprisingly indistinguishable inside the organization of ability signers.

- Unforgeability: The chances of an outside attacker successfully forging a valid signature ought to be narrow when they collect the signature of a message "m" from a random member who generates a ring signature without having access to any member's non-public key. This criterion makes it extremely unlikely to fake a legitimate signature, particularly in combative conditions.

- A ring signature setup lets the one signing decide on their own what they keep secret. This means they can make a solid, round logic involving everyone. This helps them do what's important for group signing without needing someone trusted or managing the team organizer. Ring signatures help people make secret groups and keep their transactions or messages safe without needing anyone else to watch or control them from the outside.

## 4.5. Proxy Signature: Enabling Delegated Signing

According to the discussion in reference [29], a proxy signature offers a machine in which a certified signer, known as the proxy signer, is given the authority to act on behalf of the true signer. This novel concept is predicated on solving the discrete logarithm trouble and offers a greater green framework than the repeated implementation of conventional virtual signature structures. Notably, the verifier does not want to own the overall public keys of any users aside from the true signer to complete the verification method. The final results are Enhanced computational overall performance in contrast to standard signature strategies.

In the context of blockchain, a proxy signature is a scenario wherein the authentic signer permits every other birthday celebration, known as the proxy signer, to supply their signature. Afterwards, the proxy signer can produce a legitimate signature in the area of the specific signer. The following steps are generally involved in this system:

- Initialization Process: This segment places signature device parameters further to user-precise keys.
- Power Delegation Process: A genuine signer can deliver permission to a proxy signer to join on their behalf using this process.
- Generation of a Proxy Signature: By using the unique signer's permission, the proxy signer creates a legitimate signature in this machine.
- Authentication of the Proxy Signature: The produced proxy signature is examined to verify its legitimacy.

In situations simultaneously, as delegated signing authority is needed, proxy signatures offer a useful alternative to traditional digital signatures. In the context of blockchain, Table 1 gives a contrast of 5 not unusual digital signature techniques. Because they encompass both the production of the signature and verification tactics, digital signatures—which can be based totally on public-key cryptography—are important to verifying virtual records. These signatures are an unchangeable string of numbers that attest to the integrity of the facts the sender supplied. The digital signature era is the basis of verbal exchange and is considered between nodes in a decentralized blockchain community. It permits identification verification, facts non-repudiation, and statistics integrity and authenticity assurance.

| Name of the Digital signatures | Principle | Authentication | Completeness | Integrity | Non-repudiation | Tamper Detection | Efficiency | Security | Scalability | Cross-platform Compatibility | Timestamping | Global Acceptance | Long-Term Validity | Unlinkability | Traceability | Application area |
|---|---|---|---|---|---|---|---|---|---|---|---|---|---|---|---|---|
| Aggregate signature (AS) | Co-GDH groups and bilinear mapping. | X | ✓ | X | ✓ | ✓ | ✓ | ✓ | ✓ | ✓ | X | X | ✓ | X | X | Bitcoin or Ethereum |
| Group signatures (GS) | Non-repudiation signatures | ✓ | X | ✓ | ✓ | X | ✓ | ✓ | ✓ | X | X | X | X | ✓ | ✓ | voting, document access, and privacy-preserving credentials. |
| Ring signatures (RS) | Having RSA and Labin-based versions, | X | X | ✓ | ✓ | X | ✓ | ✓ | ✓ | X | X | X | X | X | ✓ | Secure voting, and decentralized identity. |
| Blind signature (BS) | Based on RSA and DSA | ✓ | X | X | ✓ | X | X | ✓ | ✓ | X | X | X | ✓ | ✓ | ✓ | Voter privacy and ballot integrity |
| Proxy signature (PS) | Based on the discrete logarithm problem groups | ✓ | X | ✓ | X | X | ✓ | ✓ | ✓ | X | X | X | ✓ | X | X | ECC-based schemes. |

Table 1: In-depth evaluations of digital signatures utilization in blockchain

## 5. Review of related work and Comparative Analysis

In this phase, we embark on a scientific exploration of virtual signatures that discover utility in the blockchain domain. Subsequently, we provide:

- A comprehensive comparative analysis of those signature schemes.
- Delving into diverse factors, inclusive of their application fields.
- Methodologies.
- Safety attributes.
- Overall performance metrics.

This look seeks to provide a comprehensive review of the virtual signature technologies carried out within blockchain, permitting a nuanced comprehension of their strengths and weaknesses when applied in practical contexts. The proposed work of this paper is to beautify blockchain technology via virtual signatures, ensuring non-repudiation and facilitating the improvement of steady signature algorithms tailored for blockchain security and belief.

The study [30] examines the creation of a secure smart contract using blockchain technology to handle the issue of non-repudiation. With the rise of digital transactions and the need for efficient and safe contract management, the study emphasises the integration of smart contracts with blockchain to build a decentralised ledger, thereby reducing the need for central authorities. The suggested system intends to improve security by accomplishing a number of goals, including confidentiality.

This study [31] suggests that the Networking and Cryptography Library (NaCl) be enhanced to provide non-repudiation for blockchain platforms, thereby filling a key security vacuum in the library's capabilities. By analysing the NaCl library using BAN logic, the paper reveals its current inability to achieve non-repudiation, which is an essential quality for ensuring that the sender of a message cannot deny sending it. The proposed approach involves inserting a signature block into the NaCl library, which is subsequently checked for accuracy using BAN logic. This patch intends to improve blockchain system security by ensuring non-repudiation, in addition to existing features like as secrecy, integrity, and authenticity.

The paper [32] states that the emergence of quantum computing poses a threat to current cryptography standards, particularly in blockchain systems such as RSA and ECDSA. Transitioning to post-quantum cryptography (PQC) is critical for defending against quantum assaults. This move entails incorporating NIST-recommended post-quantum signature techniques into blockchain frameworks, with emphasis on speed, scalability, and security. Despite hurdles like as technical complexity and interoperability, implementing PQC improves security, promotes cryptographic innovation, and prepares blockchain networks for quantum attacks.

The study [33] emphasizes how real-time data analysis has altered farming operations. Despite its advantages, maintaining data security remains a struggle. Existing solutions frequently suffer from performance concerns and vulnerabilities. To overcome this, a unique technique is given that employs homomorphic this study is based on Hyper-Elliptic Curves. This strategy improves security while lowering computing and transmission expenses. It is validated against known attacks and provides a reliable solution for smart agriculture. Data-driven decisions that use IoT sensors to monitor soil-based and soilless agriculture optimise resource efficiency and crop output. Nonetheless, cybersecurity challenges exist. Previous research has concentrated on authentication systems, but worries regarding costs and vulnerabilities persist. The suggested approach seeks to address these issues by providing authentication, confidentiality, and threat resistance while requiring low overhead. In essence, it offers a secure and efficient solution for smart agriculture, paving the way for enhanced data protection and system resilience.

A technique for hiding transaction quantities in the privacy-focused cryptocurrency Monero was introduced in [34]. Comparable to Bitcoin,

Monero is a cryptocurrency that issues coins via a proof-of-work "mining" method and operates decentralized without a central authority. The paper [35] , [51], [52] explores the diverse applications of blockchain technology (BCT) across industries, emphasizing its transformative impact on efficiency, security, and cost reduction. It discusses real-time implementations and future use cases, categorizing business applications to aid developers and practitioners, showcasing widespread adoption and benefits ranging from financial services to healthcare, insurance, real estate, music, logistics, and government sectors, and offering improved transparency, security, and efficiency.

In [36], efforts are made to improve blockchain privacy by introducing a compatible ring signature scheme. Implemented using secp256k1, it facilitates integration into Ethereum smart contracts. The study highlights the solution's privacy and security advantages, comparing its efficiency to other privacy methods, albeit incurring costs.

The paper [37] highlights that Bitcoin transactions are publicly recorded and stored in the blockchain, necessitating secure wallet management. Losing the wallet key results in permanent Bitcoin loss, prompting the proposal of a system allowing participants to obtain a single share for key management, aligning with the weight concept's requirements.

In [38], a more streamlined ring signature scheme utilizing a compacted Non-Interactive Zero-Knowledge (NIZK) proof of knowledge is presented. This enhanced signature scheme retains its anonymity and unforgeability properties while reducing the storage space required for signatures and minimizing pairing computations during verification.

The work in [39] enhances blockchain security against potential quantum threats by employing a lattice-based signature scheme. This scheme guarantees randomness within lightweight, nondeterministic wallets, providing security within a random oracle model and contributing significantly to post-quantum blockchain research.

In reference [40], central difficulties within the domain of blockchain technology, with a particular focus on Bitcoin, are underscored. Moreover, it introduces an aggregate signature scheme, a digital signature technique that combines several signatures from various users on different messages to produce a single, streamlined signature.

The paper in [41] explains the mathematical underpinnings of virtual signatures, focusing on ECDSA, and provides insights into how ECDSA is employed in the Bitcoin environment.

Addressing the difficulty of patients managing their Electronic Health Records (EHRs) using blockchain technology, reference [42] introduces an attribute-based signature scheme via multiple governments.

The paper [43] introduces an exceptional digital proxy signature device that permits distinctive proxy signers to sign documents on behalf of the unique signer, with different delegation degrees for categorization. It provides a direct, green method, doing away with the need for verifiers to own other customers' public keys and reveals packages beyond the discrete logarithm-based total schemes.

Reference [44] explores the management of Bitcoin trading in scenarios where a Bitcoin account is mutually possessed by multiple contributors while ensuring the anonymity of those proprietors.

In [45], a centralized coin-blending algorithm known as Blind-Mixing is proposed to heighten the anonymity of Bitcoin transactions using an elliptic curve blind signature scheme.

The paper [46] describes a new approach for creating threshold ring signature schemes similar to ordinary ones. A commentary on the similarity between polynomial interpolation and the erasure correction technique of Reed-Solomon (RS) codes served as the impetus for the recently advised ring signature machine.

The paper [47] suggests a group of ring signature schemes making use of the lattice basis delegation technique. In the paper [48], the proposed work centers on improving and implementing a unique ring signature scheme tailored for blockchain structures, emphasizing compatibility, privacy, and safety.

The paper [49] addresses the vital issue of cryptographic algorithm compromise in blockchain technology and proposes a method to enhance blockchain safety by adapting a long-term signature scheme in a decentralized context, all while maintaining the integrity of the blockchain without requiring hard forks.

In reference [50], a blockchain-based LHPS (Lattice-based Hierarchical Predicate Signatures) scheme is introduced. Security analysis demonstrates that this scheme attains unforgeability even when subjected to adaptively selected message assaults, relying on the Computational Diffie-Hellman tough assumption and preserving existential unforgeability.

| Types Digital signature | Reference | Techniques used | Security | Effectiveness |
|---|---|---|---|---|
| Aggregate signature (AS) | [39] | a lattice-based signature scheme for post-quantum blockchain networks. | PoS-based consensus procedure using unconditionally secure multi-party coin flipping over QKD secured channels | Combines RandBasis and ExtBasis algorithms to enhance security, with a focus on adaptively chosen message attack resilience and efficiency, making it suitable for post-quantum blockchain networks and contributing to future PQB research. |
| | [38] | Blockchain technology, multi-signatures, and anonymous encrypted messaging streams enable secure and private energy trading with anonymity in negotiating energy prices and performing transactions | Securing decentralized smart grid energy trading without the need for trusted third parties | Token-based private decentralized energy trading system in smart grids, addressing transaction security without relying on trusted third parties. The combination of these technologies offers an effective and reliable approach to decentralized smart grid energy trading, surpassing traditional centralized solutions in terms of privacy and security. |
| | [37] | Weighted threshold signature scheme compatible with Bitcoin's elliptic curve digital signature algorithm (ECDSA) | Threshold ECDSA scheme to a weighted threshold ECDSA scheme, introducing a more granular control mechanism and enhancing security for Bitcoin transactions. | ECDSA offers significant potential for enhancing Bitcoin security, particularly in organizational contexts |
| | [40] | New signature scheme for blockchain transactions based on aggregate signatures and elliptic curve cryptography | Cryptographic scheme or system can withstand various security threats and attacks. | Combines privacy-preserving features, constant-size signatures, and big data transaction support, with potential for further enhancement through ring signatures, addressing current security gaps. |
| Group signatures (GS) | [41] | ECDSA and DSA Mathematics, ECDSA in Bitcoin Technology | Bitcoin's security is achieved through a combination of cryptographic techniques, decentralization, transparent record-keeping, and the consensus mechanism | Based on elliptic curve digital signatures and the use of the Elliptic Curve Digital Signature Algorithm (ECDSA) |
| | [42] | Multiple Authorities Attribute-Based Signature (MA-ABS) scheme to preserve patient privacy in an Electronic Health Records (EHRs) system on the blockchain. | Implementing security measures to protect patient data and the blockchain network is essential. This includes encryption, access control, and mechanisms to resist collusion attacks among authorities. | Assessed based on its ability to preserve patient privacy, ensure anonymity, guarantee data immutability, resist collusion attacks, scale with the number of authorities and patient attributes, provide formal security proofs, and integrate effectively with blockchain technology. |
| Blind signature (BS) | [43] | Digital proxy signature system with various delegation levels, based on the discrete logarithm problem | Efficient digital proxy signature scheme with various delegation levels, offering enhanced security and applicability for organizations dealing with extensive signing tasks in a digital age | Proposed partial delegation proxy signature offers computational advantages over the warrant-based version. In cases of proxy misuse, the original signer can identify the offending proxy signer and revoke the misused proxies, demonstrating the scheme's effectiveness in ensuring accountability and control. |
| | [42] | Blind signature schemes enable a recipient to obtain a signature without revealing any information about the message to the signer | The system ensures security against forgery, privacy of participants, and integrity of transaction agreements, with unlinkability and blindness properties, safeguarding Bitcoin trading. | The scheme demonstrates efficiency with practical computational complexity, making it feasible for real-world applications. |
| | [45] | Three blind signature-based algorithms: Blind-Mixing, BlindCoin, and RSA Coin-Mixing. | a Blind-Mixing system, confirming its feasibility for practical applications while ensuring enhanced security and privacy. | RSA and ECC blind signature algorithms confirming its practical feasibility while ensuring enhanced efficiency. |
| Ring Signatures (RS) | [46] | The system should seamlessly integrate with a standard public key infrastructure (PKI). | An accountable ring signature allows for verification that the signature is generated by a user from a dynamically chosen set of possible signers. | Mechanism in ring signature schemes can lead to significant consequences. It asserts that accountable ring signatures could address this issue effectively. |
| | [47] | lattice basis delegation technique | To ensure the highest level of security, which include anonymity protection against full key exposure and unforgeability in the presence of insider corruption. | cryptographic technique that allows a message to be signed by one of a group of possible signers without disclosing the specific signer's identity |
| | [48] | Compatibility with blockchain libraries and facilitates the implementation of Ethereum smart contracts. | The privacy and security aspects offered by our URS scheme | Balance between ensuring anonymity guarantees and program availability is determined by the user's preferences, it results in higher implementation costs when using Ethereum. |
| Proxy signature (PS) | [49] | blockchain technology, which encompasses Peer-to-Peer (P2P) technology, cryptography, and consensus mechanisms over a distributed network. | blockchain technology, particularly for Bitcoin and public blockchains, to ensure long-term data validity and security. | hash algorithm's impact on block size consumption is influenced by the output length of the new hash function and the volume of transactions within the block, particularly the number of interlinked transaction hash values, making it a more favourable approach compared to alternative methods. |
| | [50] | Linearly Homomorphic Proxy Signature (LHPS) schemes for data security in the context of outsourcing computing tasks in IoT | LHPS scheme enhances security by achieving unforgeability chosen message attacks, reducing it to the CDH hard assumption, and ensuring both usual and homomorphic existential unforgeability, while maintaining key size and computational efficiency. | Correctness of computation results, and reliability in IoT outsourcing scenarios while maintaining key size and computational efficiency. |

Table 2: Comparing Digital Signature Schemes: An Evaluation and Analysis

## 6. Result and Discussion

This study analyses various digital signature methods and analyses their performance on a variety of factors, including security, computational efficiency, and application applicability. The Elliptic Curve Digital Signature Algorithm (ECDSA) stands out for its excellent security per key length, faster processing rates, smaller key sizes, and lower bandwidth needs. For example, a 160-bit ECDSA key provides the same level of security as a 1024-bit RSA key but uses far less processing power and storage space.

Our studies discovered limits in conventional digital signature methods such as RSA and DSA, which display performance inefficiencies with rising key lengths, limiting their appropriateness for blockchain encryption/decryption operations. ECDSA improves efficiency, although it may struggle in resource-constrained IoT contexts. In order to resolve these difficulties, continuing research is aimed at creating more efficient lattice-based and aggregate signature techniques. These developing ideas seek to address privacy and performance concerns in blockchain transactions by improving the scalability and security of digital signatures. Such developments open the path for more secure, efficient, as well as scalable digital signature implementations, resulting in strong and trustworthy digital systems.

## 7. Conclusion and future work

The study analyzed digital signature methods, highlighting ECDSA's superior security, efficiency, and scalability for blockchain applications compared to RSA and DSA. It identified limitations in conventional methods and proposed ongoing research into lattice-based and aggregate signatures to address privacy and performance concerns. This research paves the way for more secure, efficient, and scalable digital signature implementations, enhancing the trustworthiness of blockchain systems. Improving information security and non-repudiation in blockchain structures has been the main consciousness of this research. A comparative study of the numerous digital signature schemes used within cutting-edge blockchain applications proves these technologies can meet protection requirements in plenty of eventualities and efficiently address precise blockchain necessities. The contributions of this paper are meant to assist security by presenting recommendations for developing and enhancing blockchain-primarily based digital signature schemes.

It is usually recommended that the subsequent moves be taken to improve security and pastime in this area:

1. Improving Transaction Anonymity: When it involves Bitcoin transactions on the blockchain, consider using techniques like ring signatures and blind signatures to analyze and decorate the diploma of transaction information encryption and participant anonymity to lower transaction overhead ultimately.

2. Multidimensional Security: To toughen protection from several guidelines and guarantee sturdy non-repudiation of facts within the blockchain, inspect using digital signatures and identity authentication or timestamping.

3. Blockchain and Internet of Things Integration: Since the two are increasingly combined, focus on creating certificate aggregate signature techniques, particularly for IoT devices with constrained assets, especially mobile terminals, to handle their particular safety problems.

Future work involves exploring more efficient lattice-based and aggregate signature techniques to address the limitations of conventional digital signature methods like RSA and DSA. These efforts aim to enhance the scalability and security of digital signatures in blockchain transactions, paving the way for stronger, more efficient, and scalable digital systems.

**Reference**

[1] Zhou, L., Zhang, L., Zhao, Y., Zheng, R., & Song, K. (2021). A scientometric review of blockchain research. Information Systems and e-Business Management, 1-31. https://doi.org/10.1007/s10257-020-00461-9

[2] Baiod, W., Light, J., & Mahanti, A. (2021). Blockchain technology and its applications across multiple domains: A survey. Journal of International Technology and Information Management, 29(4), 78-119. https://doi.org/10.58729/1941-6679.1482

[3] Zwitter, A., & Hazenberg, J. (2020). Decentralized network governance: blockchain technology and the future of regulation. Frontiers in Blockchain, 3, 12. https://doi.org/10.3389/fbloc.2020.00012

[4] Stockburger, L., Kokosioulis, G., Mukkamala, A., Mukkamala, R. R., & Avital, M. (2021). Blockchain-enabled decentralized identity management: The case of self-sovereign identity in public transportation. Blockchain: Research and Applications, 2(2), 100014. https://doi.org/10.1016/j.bcra.2021.100014

[5] Udokwu, C., Anyanka, H., & Norta, A. (2020, January). Evaluation of approaches for designing and developing decentralized applications on blockchain. In Proceedings of the 4th International Conference on Algorithms, Computing and Systems (pp. 55-62). https://doi.org/10.1145/3423390.3426724

[6] Leng, J., Ruan, G., Jiang, P., Xu, K., Liu, Q., Zhou, X., & Liu, C. (2020). Blockchain-empowered sustainable manufacturing and product lifecycle management in industry 4.0: A survey. Renewable and sustainable energy reviews, 132, 110112. https://doi.org/10.1016/j.rser.2020.110112

[7] Khezami, N., Gharbi, N., Neji, B., & Braiek, N. B. (2022). Blockchain Technology Implementation in the Energy Sector: Comprehensive Literature Review and Mapping. Sustainability, 14(23), 15826. https://doi.org/10.3390/su142315826

[8] Dai, Y., Xu, D., Zhang, K., Maharjan, S., & Zhang, Y. (2020). Deep reinforcement learning and permissioned blockchain for content caching in vehicular edge computing and networks. IEEE Transactions on Vehicular Technology, 69(4), 4312-4324. https://doi.org/10.1109/TVT.2020.2973705

[9] Javaid, M., Haleem, A., Singh, R. P., Khan, S., & Suman, R. (2021). Blockchain technology applications for Industry 4.0: A literature-based review. Blockchain: Research and Applications, 2(4), 100027. https://doi.org/10.1016/j.bcra.2021.100027

[10] Hashemi Joo, M., Nishikawa, Y., & Dandapani, K. (2020). Cryptocurrency, a successful application of blockchain technology. Managerial Finance, 46(6), 715-733. https://doi.org/10.1108/MF-09-2018-0451

[11] Fauziah, Z., Latifah, H., Omar, X., Khoirunisa, A., & Millah, S. (2020). Application of blockchain technology in smart contracts: A systematic literature review. Aptisi Transactions on Technopreneurship (ATT), 2(2), 160-166. https://doi.org/10.34306/att.v2i2.97

[12] Wenhua, Z., Qamar, F., Abdali, T. A. N., Hassan, R., Jafri, S. T. A., & Nguyen, Q. N. (2023). Blockchain technology: security issues, healthcare applications, challenges and future trends. Electronics, 12(3), 546. https://doi.org/10.3390/electronics12030546

[13] Guru, A., Mohanta, B. K., Mohapatra, H., Al-Turjman, F., Altrjman, C., & Yadav, A. (2023). A survey on consensus protocols and attacks on blockchain technology. Applied Sciences, 13(4), 2604. https://doi.org/10.3390/app13042604

[14] Schlatt, V., Guggenberger, T., Schmid, J., & Urbach, N. (2023). Attacking the trust machine: Developing an information systems research agenda for blockchain cybersecurity. International journal of information management, 68, 102470. https://doi.org/10.1016/j.ijinfomgt.2022.102470

[15] Alanzi, H., & Alkhatib, M. (2022). Towards Improving Privacy and Security of Identity Management Systems Using Blockchain Technology: A Systematic Review. Applied Sciences, 12(23), 12415. https://doi.org/10.3390/app122312415

[16] Gururaj, H. L., Manoj Athreya, A., Kumar, A. A., Holla, A. M., Nagarajath, S. M., & Ravi Kumar, V. (2020). Blockchain: A new era of technology. Cryptocurrencies and blockchain technology applications, 1-24. https://doi.org/10.1002/9781119621201.ch1

[17] Lv, Z., Qiao, L., Hossain, M. S., & Choi, B. J. (2021). Analysis of using blockchain to protect the privacy of drone big data. IEEE network, 35(1), 44-49. https://doi.org/10.1109/MNET.011.2000154

[18] Zheng, Z., Xie, S., Dai, H. N., Chen, X., & Wang, H. (2018). Blockchain challenges and opportunities: A survey. International journal of web and grid services, 14(4), 352-375. https://doi.org/10.1504/IJWGS.2018.095647

[19] Dai, F., Shi, Y., Meng, N., Wei, L., & Ye, Z. (2017, November). From Bitcoin to cybersecurity: A comparative study of blockchain application and security issues. In 2017 4th International Conference on Systems and Informatics (ICSAI) (pp. 975-979). IEEE. https://doi.org/10.1109/ICSAI.2017.8248427

[20] Sikeridis, D., Bidram, A., Devetsikiotis, M., & Reno, M. J. (2020, January). A blockchain-based mechanism for secure data exchange in smart grid protection systems. In 2020 IEEE 17th Annual Consumer Communications & Networking Conference (CCNC) (pp. 1-6). IEEE. https://doi.org/10.1109/CCNC46108.2020.9045368

[21] Mathur, S., Kalla, A., Gür, G., Bohra, M. K., & Liyanage, M. (2023). A Survey on Role of Blockchain for IoT: Applications and Technical Aspects. Computer Networks, 227, 109726. https://doi.org/10.1016/j.comnet.2023.109726

[22] Zieglmeier, V., Loyola Daiqui, G., & Pretschner, A. (2023). Decentralized inverse transparency with blockchain. Distributed Ledger Technologies: Research and Practice, 2(3), 1-28. https://doi.org/10.1145/3592624

[23] Liu, S. G., Chen, W. Q., & Liu, J. L. (2021). An efficient double parameter elliptic curve digital signature algorithm for blockchain. IEEE Access, 9, 77058-77066. https://doi.org/10.1109/ACCESS.2021.3082704

[24] Bralić, V., Stančić, H., & Stengård, M. (2020). A blockchain approach to digital archiving: digital signature certification chain preservation. Records Management Journal, 30(3), 345-362. https://doi.org/10.1108/rmj-08-2019-0043

[25] Capece, G., Levialdi Ghiron, N., & Pasquale, F. (2020). Blockchain technology: Redefining trust for digital certificates. Sustainability, 12(21), 8952. https://doi.org/10.3390/su12218952

[26] Li, G., & Ou, C. (2023, October). Blockchain Encryption Algorithm Based on Aggregated Signature. In 2023 IEEE 3rd International Conference on Data Science and Computer Application (ICDSCA) (pp. 1011-1016). IEEE. https://doi.org/10.1109/ICDSCA59871.2023.10392603

[27] Li, C., Tian, Y., Chen, X., & Li, J. (2021). An efficient anti-quantum lattice-based blind signature for blockchain-enabled systems. Information Sciences, 546, 253-264. https://doi.org/10.1016/j.ins.2020.08.032

[28] Li, X., Mei, Y., Gong, J., Xiang, F., & Sun, Z. (2020). A blockchain privacy protection scheme based on ring signature. IEEE Access, 8, 76765-76772. https://doi.org/10.1109/ACCESS.2020.2987831

[29] Wang, Y., Qiu, W., Dong, L., Zhou, W., Pei, Y., Yang, L., ... & Lin, Z. (2020). Proxy signature-based management model of sharing energy storage in blockchain environment. Applied Sciences, 10(21), 7502. https://doi.org/10.3390/app10217502

[30] Mohammed, N. S., Dawood, O. A., Sagheer, A. M., & Nafea, A. A. (2024). Secure Smart Contract Based on Blockchain to Prevent the Non-Repudiation Phenomenon. Baghdad Science Journal, 21(1), 0234-0234. https://doi.org/10.21123/bsj.2023.8164

[31] Rasslan, M., Nasreldin, M. M., Abdelrahman, D., Elshobaky, A., & Aslan, H. (2024). Networking and cryptography library with a non-repudiation flavor for blockchain. Journal of Computer Virology and Hacking Techniques, 20(1), 1-14. https://doi.org/10.1007/s11416-023-00482-1

[32] Javaid, M. A. R., Ashraf, M., Rehman, T., & Tariq, N. (2024). Impact of Post Quantum Digital Signatures On Block Chain: Comparative Analysis. The Asian Bulletin of Big Data Management, 4(1), Science-4. https://doi.org/10.62019/abbdm.v4i1.133

[33] TAJI, K., & GHANIMI, F. (2024). Enhancing Security and Privacy in Smart Agriculture: A Novel Homomorphic Signcryption System. Results in Engineering, 102310. https://doi.org/10.1016/j.rineng.2024.102310

[34] Chithaluru, P., Singh, K., & Sharma, M. K. (2020). Cryptocurrency and blockchain. Information security and optimization, 143-158. https://doi.org/10.1201/9781003045854

[35] Laroiya, C., Saxena, D., & Komalavalli, C. (2020). Applications of blockchain technology. In Handbook of research on blockchain technology (pp. 213-243). Academic press. https://doi.org/10.1016/B978-0-12-819816-2.00009-5

[36] Ferrag, M. A., & Shu, L. (2021). The performance evaluation of blockchain-based security and privacy systems for the Internet of Things: A tutorial. IEEE Internet of Things Journal, 8(24), 17236-17260. https://doi.org/10.1109/JIOT.2021.3078072

[37] Dikshit, P., & Singh, K. (2017). Weighted threshold ECDSA for securing bitcoin wallet. In 2017 ISEA Asia Security and Privacy (ISEASP). http://dx.doi.org/10.19101/TIS.2017.26003

[38] Ren, H., Zhang, P., Shentu, Q., Liu, J. K., & Yuen, T. H. (2018). Compact ring signature in the standard model for blockchain. In Information Security Practice and Experience: 14th International Conference, ISPEC 2018, Tokyo, Japan, September 25-27, 2018, Proceedings 14 (pp. 50-65). Springer International Publishing. http://dx.doi.org/10.1007/978-3-319-99807-7_4

[39] Li, C. Y., Chen, X. B., Chen, Y. L., Hou, Y. Y., & Li, J. (2018). A new lattice-based signature scheme in post-quantum blockchain network. IEEE Access, 7, 2026-2033. http://dx.doi.org/10.1109/ACCESS.2018.2886554

[40] Yuan, C., Xu, M. X., & Si, X. M. (2017). Research on a new signature scheme on blockchain. Security and Communication Networks, 2017. https://doi.org/10.1155/2017/4746586

[41] Kikwai, B. K. (2017). Elliptic curve digital signatures and their application in the bitcoin crypto-currency transactions.

[42] Guo, R., Shi, H., Zhao, Q., & Zheng, D. (2018). Secure attribute-based signature scheme with multiple authorities for blockchain in electronic health records systems. IEEE access, 6, 11676-11686. https://doi.org/10.1109/ACCESS.2018.2801266

[43] Mambo, M., Usuda, K., & Okamoto, E. (1996, January). Proxy signatures for delegating signing operation. In Proceedings of the 3rd ACM Conference on Computer and Communications Security (pp. 48-57). https://doi.org/10.1145/238168.238185

[44] Wu, Q., Zhou, X., Qin, B., Hu, J., Liu, J., & Ding, Y. (2017). Secure joint bitcoin trading with partially blind fuzzy signatures. Soft Computing, 21, 3123-3134. https://doi.org/10.1007/s00500-015-1997-6

[45] ShenTu, Q., & Yu, J. (2015). A blind-mixing scheme for bitcoin based on an elliptic curve cryptography blind digital signature algorithm. arXiv preprint arXiv:1510.05833. https://doi.org/10.48550/arXiv.1510.05833

[46] Xu, S., & Yung, M. (2004). Accountable ring signatures: A smart card approach. In Smart Card Research and Advanced Applications VI: IFIP 18th World Computer Congress TC8/WG8. 8 & TC11/WG11. 2 Sixth International Conference on Smart Card Research and Advanced Applications (CARDIS) 22–27 August 2004 Toulouse, France (pp. 271-286). Springer US. https://doi.org/10.1007/1-4020-8147-2_18

[47] Wang, J., & Sun, B. (2011). Ring signature schemes from lattice basis delegation. In Information and Communications Security: 13th International Conference, ICICS 2011, Beijing, China, November 23-26, 2011. Proceedings 13 (pp. 15-28). Springer Berlin Heidelberg. https://doi.org/10.1007/978-3-642-25243-3_2

[48] Mercer, R. (2016). Privacy on the blockchain: Unique ring signatures. arXiv preprint arXiv:1612.01188. https://doi.org/10.48550/arXiv.1612.01188

[49] Sato, M., & Matsuo, S. I. (2017, July). Long-term public blockchain: Resilience against compromise of underlying cryptography. In 2017 26th International Conference on Computer Communication and Networks (ICCCN) (pp. 1-8). IEEE. https://doi.org/10.1109/ICCCN.2017.8038516

[50] Wang, C., & Wu, B. (2023). A Linear Homomorphic Proxy Signature Scheme Based on Blockchain for Internet of Things. CMES-Computer Modeling in Engineering & Science,136(2). https://doi.org/10.32604/cmes.2023.026153

[50] Sajal Saha et al., "THMIP- A Novel Mobility Management Scheme using Fluid Flow model", 2nd International Conference on Emerging Trends and Applications in Computer Science, (NCETACS - 2011), India, 4th- 6th March, 2011

[51] S. Khara, Sajal Saha, A. K Mukhopadhyay, C. Ghosh," Call Dropping Analysis in a UMTS/WLAN Integrated Cell", p.p. 411-415, International Journal of IT and Knowledge Management (IJITKM) (ISSN 0973-4414), Vol-III, Issue-II.